\documentclass[superscriptaddress,twocolumn,showpacs]{revtex4}
\usepackage{graphicx}
\usepackage{dcolumn}
\usepackage{bm}

\newcommand{\bb}{\begin{equation}}
\newcommand{\en}{\end{equation}}

\begin{document}

\title{Dynamics of bacterial flow: Emergence of spatiotemporal coherent
structures}
\author{Nicholas Sambelashvili}
\affiliation{Department of Physics, New York University, New York, NY 10012, USA}
\author{A.W.C. Lau}
\affiliation{Department of Physics, Florida Atlantic University, Boca Raton, FL 33431, USA}
\author{David Cai}
\affiliation{Courant Institute of Mathematical Sciences, New York University, New York,
NY 10012, USA}
\date{\today}

\begin{abstract}
We propose a simple model of self-propelled particles to show that coherent
structures, such as jets and swirls, can arise from a plausible microscopic
mechanisms: (i) the elongated shape of the self-propelled particles with
(ii) the hardcore interactions among them. We demonstrate via computer
simulation that these coherent structures, which emerge at sufficiently high
densities of particles, have characteristics that are similar to those
observed in recent experiments in bacteria baths.
\end{abstract}

\maketitle

Living systems often exhibit complex spatiotemporal patterns that are
characteristic of many systems driven out-of-equilibrium \cite{gollub}.
Examples range from birds flocking \cite{vicsek} to internal organizations
inside a cell \cite{nedelec}. A population of live microorganisms, such as
bacteria like E.\ coli, suspended in an aqueous environment can provide a
particular interesting model system to study pattern formation and
collective motion in biology since the dynamics of the individual bacteria
can be directly observed and critical parameters such as density and
activity may be brought under experimental control \cite{berg}. These
microorganisms continuously consume nutrients and dissipate the energy
through the process of propelling themselves against the frictional force
exerted on them by the fluid \cite{purcell}. With a typical size of a
bacterium of the order of microns and a typical speed of the order of 10 $%
\mu $m/s, the Reynolds number $R \ll 1$ is quite small. Yet, a large
concentration of these microorganisms constitutes a state that is far from
equilibrium, and exhibits self-organized collective motion with spatial and
temporal patterns that are both physically fascinating and potentially of
great biological significance \cite{mendelson,wu,soni,goldstein}. As a first
step towards understanding of how suspended cells generate coherent motion,
we identify, in this Letter, two simple but central ingredients -- the
elongated shape of the self-propelled particles and the hardcore
interactions among them, and demonstrate via computer implementation of
these two ingredients that this system exhibits coherent jets and swirls
with characteristics strikingly similar to those observed in experiments.

Recent experiments \cite{mendelson,wu,soni,goldstein} show that when
concentrated, the crowd of swimming bacteria creates arrays of transient
jets and swirls whose size can be orders of magnitude larger than an
individual bacterium. These complicated, spatially coherent structures have
been observed in Bacillus subtilis colony grown on an agar plate \cite%
{mendelson}, in E. coli confined in a quasi-two-dimensional soap film \cite%
{wu}, in Bacillus subtilis at the edge of a pendent drop \cite{goldstein}.
It is estimated from direct visualization that these structures have a
typical size of about ten times that of a bacterium and persist for a few
seconds. More quantitative information about these structures may be
extracted from a one-point passive microrheological technique which tracks
passive micron-sized beads dispersed in a bacterial bath of E.\ coli \cite%
{wu}. Interestingly, the mean-squared displacement (MSD) of these passive
beads exhibit superdiffusion at short time and diffusion at long time. It is
measured that the crossover time $\tau _{c}\sim 2\,$s and the length $\ell
_{c}\equiv \sqrt{\langle \lbrack \Delta x(\tau _{c})]^{2}\rangle }\sim
10\,\mu $m provide a natural time and length scales of these coherent
structures, respectively.

\begin{figure*}[tbp]
\rotatebox{0}{\resizebox*{6.0in}{!}{\includegraphics{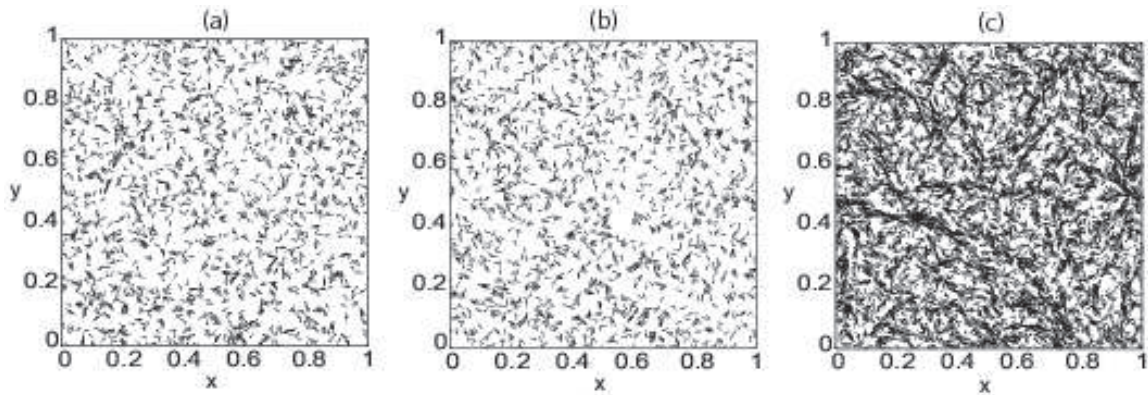}}} %
\caption{Snapshots of bacterial motion at low and intermediate particle
density. (a) and (b): The number of particles is $N=2000$. These two
snapshots are one second apart. Each bacterium is represented as a vector
whose length is the same as the bacterial length. There is no apparent
coherent structure present. The bacteria move as if they were statistically
independent, although some groupings occur accidentally, but they disappear
quickly. The distribution is quite homogeneous and isotropic.
(c) $N=8000$ --- Emergence of coherent structures; note the dark stripes
where a bunch of bacteria moves in similar directions.}
\label{2000}
\end{figure*}

There have been a few simulation models and theoretical treatments that aim
to describe the above and related phenomena \cite{vicsek,tu,tu1,ramaswamy}.
Motivated by the patterns in fish schooling and bird flocking, Vicsek \emph{%
et al.\ } numerically studied a model in which each particle (modelled as a
point) moves at a constant speed and its direction is determined by
averaging over the directions of a large collection of particles in its
neighborhood plus a small random perturbation \cite{vicsek}. This model in
2D exhibits a nonequilibrium phase transition: At sufficiently high
concentration, all particles spontaneously move in a single direction, thus
breaking the rotational symmetry of the system. This phase transition is
rationalized within a phenomenological dynamical xy-model \cite{tu}. This
model further predicts that in 2D, the MSD of a tag particle exhibits
superdiffusion at the transition. Indeed, via computer simulation of the
bird flocking model, it was shown in Ref.\ \cite{tu1} that a passive bead,
interacting with a sea of \textquotedblleft birds", exhibit superdiffusion
as observed in Ref.\ \cite{wu}. However, the coherent structures appearing
in the ordered state of the bird flocking model have a size scaled as the
system size, whereas those observed in experiments have finite size. Note
that the bird flocking model ignores the nematic-like ordering arising from
the rod-like shape of a bacterium. More recently, a novel phenomenological
macroscopic equations which take the ambient fluid and nematic ordering into
account have been proposed to describe the dynamics of suspensions of
self-propelled particles \cite{ramaswamy}. This theory is a generalization
of the equilibrium hydrodynamics of nematogens to a non-equilibrium
situation. It predicts that an ordered suspension of self-propelled nematics
is unstable at long wavelength, possibly giving rise to vortices and jets.
Although such macroscopic equations are well suited for probing structures
at large scales, it seems rather difficult to apply them to determine
microscopic mechanisms for the coherent structures down to scales of a few
particle sizes.

In this letter, we propose a new 2-D microscopic model that is based on two
simple physical ingredients: First, since the ratio of the length of a
bacterium to its width is typically $\sim 5$-$10$, we model each bacterium
as an infinitely thin rod with a fixed length. Secondly, we impose hard-core
interaction among them, \emph{i.e.,} they cannot intersect when they are
moving. Thus, our model may be viewed as a nonequilibrium generalization of
the two-dimensional hard-core nematics studied in Ref.\ \cite{needle}. As we
demonstrate below, our model in 2D exhibits coherent structures which emerge
at sufficiently high densities, even though the bacteria in our model have
no \textquotedblleft will", in the sense that they do not seek out at what
directions their neighbors are swimming, they do not communicate with each
other via chemical signaling, nor do they interact hydrodynamically. Rather,
the simple physical picture emerging from this study is that, with the
constraints of the elongated shape of the needle and the excluded volume
interaction, a self-propelled bacterium can find a new location only where
it can fit into. This forces the needles to align locally and swim together
in the form of swirls and jets. Our simulation shows that (i) typical sizes
of these coherent structures are consistent with experimental observations,
and (ii) the density fluctuations are characterized by non-Poisson
statistics.

\begin{figure*}[tbp]
\rotatebox{0}{\resizebox*{6.0in}{!}{\includegraphics{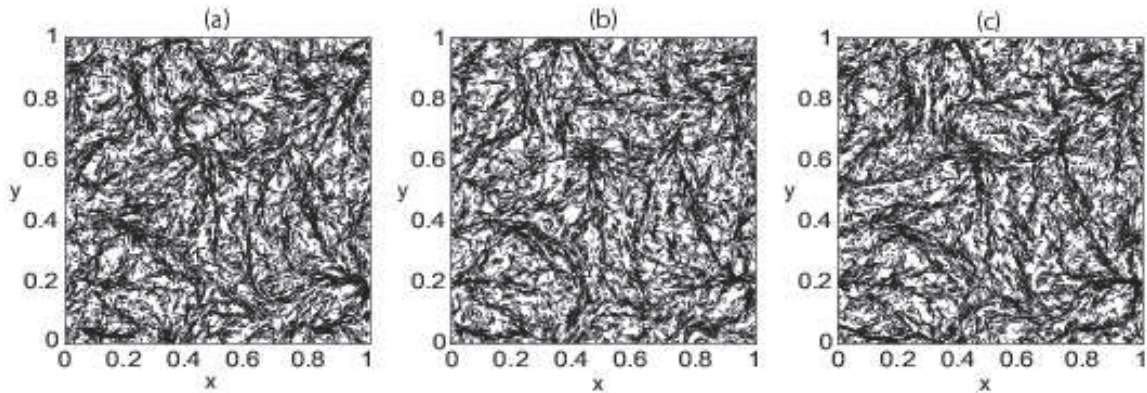}}} %
\caption{Snapshots of bacterial motion at the high particle density ($%
N=12000 $). See Fig. (\protect\ref{2000}) for the graphic representation
convention. Dense needle regions form coherent structures of jets and
swirls.  Some of these structures can be seen to persist for seconds as shown
here. (The time between each snapshot here is $1$ s.) Note the temporal
evolution of the shape and size of these structures.}
\label{12000}
\end{figure*}

\begin{figure*}[tbp]
\rotatebox{0}{\resizebox*{6.0in}{!}{\includegraphics{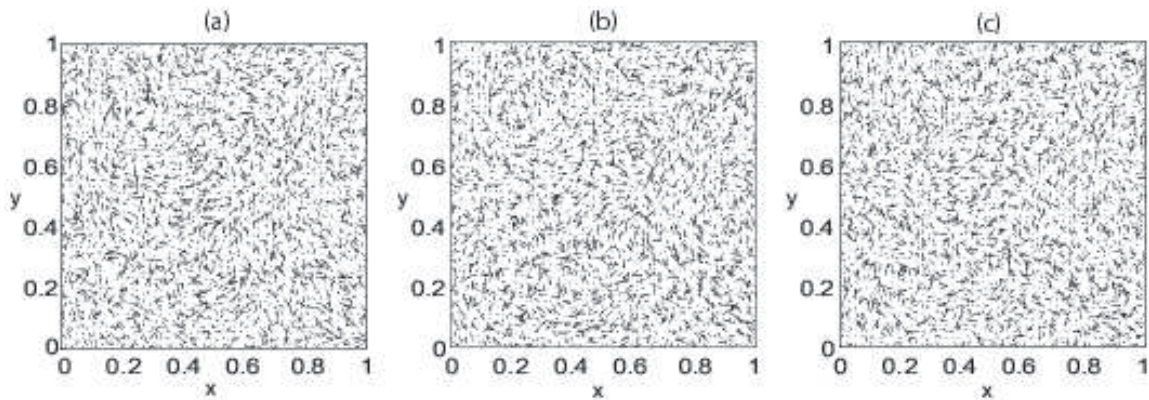}}} %
\caption{Snapshots of velocity fields, $N=12000$, (see text). Long jets and
large swirls are persistent. The size of jets is comparable to the box size;
swirls can have long life-time (around $8-10$ seconds). The time between
each snapshot here is 1 second with a total duration of two seconds shown.)}
\label{12000a}
\end{figure*}

We now detail our model of bacterial bath and describe briefly our
simulation method. We first deposit $N$ needles of length $\ell $, whose
center-of-mass coordinates $\mathbf{r}_{i}=(x_{i},y_{i})$, $i=1,\ldots ,N$,
are drawn at random from a uniform distribution in a 2D box with the length $%
L$. The orientation of each needle $\hat{\mathbf{n}}_{i}$ can be
characterized by the angle with respect to the $y$-axis. This angle is
chosen at random in the range of $[0,2\pi ]$ from a uniform distribution. At
this stage, there may be many bacteria intersecting with each other. To
impose the hard-core interactions, we must first obtain a non-intersecting
state. This is accomplished by moving each needles by a certain distance and
rotating its direction until no needles are intersecting. This state
constitutes an acceptable initial state. We use the criterion in Ref.\ \cite%
{needle} to test whether two needles are intersecting: Two needles $i$ and $j
$ intersect if and only if both $g_{i}$ and $g_{j}$ are negative, where the
quantities $g_{i}$ and $g_{j}$ are defined as%
\begin{eqnarray}
g_{i} &\equiv &(\mathbf{r}_{ij}\cdot \mathbf{v}_{i})^{2}-(\ell /2)^{2}\left[
1-(\mathbf{v}_{i}\cdot \mathbf{v}_{j})^{2}\right] , \\
g_{j} &\equiv &(\mathbf{r}_{ij}\cdot \mathbf{v}_{j})^{2}-(\ell /2)^{2}\left[
1-(\mathbf{v}_{i}\cdot \mathbf{v}_{j})^{2}\right] ,
\end{eqnarray}%
with $\mathbf{r}_{ij}=\mathbf{r}_{i}-\mathbf{r}_{j}$ and $\mathbf{v}_{i}$
and $\mathbf{v}_{j}$ being unit vectors perpendicular to $\hat{\mathbf{n}}%
_{i}$ and $\hat{\mathbf{n}}_{j}$, respectively. We employ periodic boundary
condition in both directions. The bacterial motion is simulated with the
following dynamics: a particle is chosen to move to a new position while
remaining particles are fixed. Note that this asynchronous updating avoids
complication associated with many-body collisions, which would be difficult
to resolve computationally for the hardcore interactions. Thus, at each time
step, an individual bacterium moves by a fixed distance $d$ and
simultaneously its direction changes by a small angle $\theta $, which is
chosen uniformly at random from the interval $[-\eta /2,\eta /2]$. If the
chosen bacterium intersects with another one, in order to satisfy the
hard-core constraint, we immediately discard the move and start a new trial
move: we pull it back to the previous (original) position, and change its
direction by another random angle $\theta $. If the number of such trials
exceeds a preset number $N_{\text{trial}}\sim 40$ \cite{note}, then we
increase $\eta $ by a small amount to a new $\eta ^{\prime }=a\,\eta ,$ $
\left(\text{say, }a=1.02\right) $ and repeat the trial until
this particular particle can make its move. Then we reset $\eta $
to its default value and update the next particle. We emphasize
that there is no preferential searching involved in a sense that
the bacterium is merely trying to find a new location where it can
fit into. This procedure to update the position and direction of
the needle is repeated for all particles sequentially at each time
step. We carry out our simulations, and discard the first $\sim
$3000 moves as transient. For all the simulations reported below,
we have chosen $\ell =0.04$, $L=1$, $d=0.001$, and $N$ ranges from
$\sim 10^{3}$ to $\sim 10^{4}$. The parameter $\eta $ determines
how widely a bacterium can change its direction, and its default value
is $\eta =0.01.$ In order to compare with physical values, each time
step corresponds to $\sim 0.01\,$s. Roughly, this equivalently
corresponds to the situation where a bacterium of length of a few
microns swims at a typical speed of a few $\mu \mbox{m}/\mbox{s}$
\cite{goldstein}.

We now turn to our main simulation results. Depicted in Figs.\ \ref{2000},
and \ref{12000} are snapshots of the needle configuration for $N=2000$, $8000
$, and $12000$, respectively. As evident in these figures, different
densities of particles result in different structures in their collective
motion. For low densities, around $N\sim 10^{3}$ there are no
spatiotemporally coherent structures, as shown in Fig.\ \ref{2000}(a) and
(b). Bacteria rarely collide. However, as the concentration of the particles
reaches the values of $N\sim 4000$, some coherent structures start to emerge
with a short persistent time ($t\ll 1$s). At $N=8000$, these coherent
structures become more prominent and one can easily identify long jets and
small swirls (Fig.\ \ref{2000}(c)). These structures can persist for a few
seconds. For similar densities, the same structures are observed
experimentally \cite{wu,goldstein}. As the particle concentration increases,
the typical size of the coherent structures and the frequency of their
occurrence also increase. The highest density we consider is $N=12000$. In
this case, jets of the size of the box appear and almost all particles are
involved in collective motion [Fig.\ \ref{12000}]. The evolution of the
velocity field is shown in Fig. \ref{12000a}, where the velocity field is
the average vector by summing bacterial vectors locally with the averaging
area being $\ell ^{2}$. Figure \ref{12000a} clearly demonstrates the
persistence of coherent structures for seconds in the velocity field. At a
fixed particle concentration, the size of swirls clearly depends on $\eta $.
As $\eta $ increases, the typical size of swirls decreases. This is because
particles in the swirls at each time step explore wider angles and are able
to escape from the region of the swirl. For $\eta \sim 0.01$, the size of
swirls is about $5$--$7$ times the lengths of the bacterium at $N\sim 4000\,$%
--$\,8000$. One of the temporal characteristics of jets or swirls is their
persistence time. Experimental observations \cite{goldstein} indicate that
the persistence time for \textit{swirls} is on the order of several seconds,
while jets can persist for longer time. Our simulations yield similar
results for swirls and jets. Incidentally, for initial conditions with all
parallel needles, our simulations show that this initial long-ranged order
is quickly destroyed but the coherent structures of jets and swirls again
develop as the system evolves. We emphasize that the hard-core interaction
among elongated self-propelled objects underlies all these structures in the
simulation. The physical picture emerges from this model is that the
excluded-volume interactions effectively reduces $\eta $, which forces
particles to align locally. This mechanism appears to be consistent with
experimental observations in that bacteria tumble (hence, changing
directions) less when concentrated \cite{wu}.

\begin{figure}[tbp]
{}
\par
\centering
\resizebox*{2.2in}{!}{\rotatebox{0}{\includegraphics{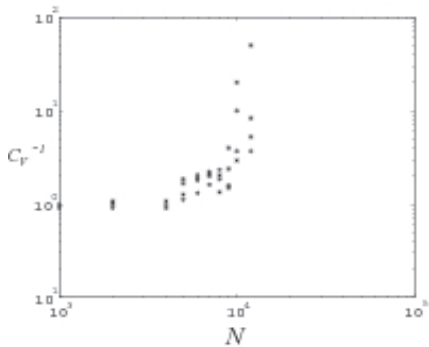}}}
\caption{Particle density fluctuation characterized by $C_V^{-1}$. $%
T_{C_V}=0.1\,$ s. For a fixed $N$, we plot here the value of $C_V^{-1}$ for
each independent run. For example, at $N=8000$, there were three runs, thus
three different values of $C_V^{-1}$. (Instead of averaging over these runs,
we display this way to indicate the statistical spread.) }
\label{CV}
\end{figure}

To further quantify the transition from the homogeneous, isotropic flow to
highly organized coherent motions of bacteria, we measure the coefficient of
variance, $C_{V}$, from our simulation to describe the coherence of the
motion, where $C_{V}=\sigma _{n}^{2}/\langle n\rangle $, $\langle n\rangle $
is the average number of bacteria appearing in a small square with an area
1/16 over a fixed time interval $T_{C_{V}}$, and $\sigma _{n}$ is the number
fluctuation over $T_{C_{V}}$. Figure \ref{CV} displays that $C_{V}^{-1}$ is
nearly one at the low density limit (Note that $C_{V}=1$ for a Poisson
statistics.). Then, $C_{V}^{-1}$ has a rapid rise around $N\sim 10^{4}$,
signifying a drastic reduction of fluctuations in the particle density.
Hence, the onset of the coherent motion of the bacterial flow.

In conclusion, we have proposed a possible mechanism for the emergence of
coherent structures in a bacterial flow: they arise in $2D$ from their
elongated shapes with the hard-core interaction. We have shown that the
model indeed gives rise to persistence of jets and swirls at sufficiently
high density of needles, with comparable spatiotemporal scales as observed
in experiments for bacteria at similar densities. When the bacterial
concentration becomes sufficiently high, one may suspect that hydrodynamic
interactions of solution become important. As noted before, our simple model
does not taken into account hydrodynamic interactions. In general, inclusion
of hydrodynamic interactions for \emph{elongated} objects is computationally
involved. However, we have performed a simulation of \textit{spherical}
particle motion following method in Ref.\ \cite{ball} inside a $3D$ unit
box, taking into account of the \emph{lubrication} effect only. In this
setup, we did not observe any jets or swirls, which means that hydrodynamics
alone probably cannot organize spherical particles into large-scale coherent
motions. This is indicative of the fact that the elongated shape together
with hardcore interaction in 2D plays a primary role for the emergence of
coherent structures even in the presence of hydrodynamics. We believe that
at high densities the formation of the coherent structures due to our
mechanism is robust qualitatively, with possible quantitative modifications
by details of hydrodynamic interactions.

We acknowledge fruitful discussion with Daniel T.\ Chen, T.C.\ Lubensky.
This work was supported by the NSF through the MRSEC Grant DMR-0079909 for
A.W.C.L., DMS-0206679 and DMS-0507901 for N.S and D.C.


\begin{thebibliography}{99}
\bibitem{note} Note that we have chosen $N_{\mbox{\scriptsize trials}} \sim
40$ in order to have as few particles as possible to be stuck in some
region. Typically, the fraction of stuck particles does not exceed $10\,\%$.
\end{thebibliography}
\end{document}